\documentclass[preprint]{aastex}
\usepackage[]{epsfig}
\newcommand       \beq          {\begin{equation}}
\newcommand       \eeq          {\end{equation}}
\newcommand       \Angstrom     {\,{\rm \AA}}          

\newcommand       \cm           {\,{\rm cm}}

\newcommand       \simlt        {\lesssim}
\newcommand       \simgt        {\gtrsim}
\newcommand       \gtsim        {\gtrsim}
\newcommand       \ltsim        {\lesssim}
\newcommand       \fsi          {{\rm Y_{\rm si}}}

\newcommand{\figwidth}{11cm}
\newcommand{\figwidthland}{12cm}
\countdef\decade=200
\advance\decade by \year
\countdef\hours=201
\advance\hours by \time
\divide\hours by 60
\countdef\mins=202
\advance\mins by \hours
\multiply\mins by 60
\multiply\hours by 100
\countdef\miltime=203
\advance\miltime by \hours
\advance\miltime by \time
\advance\miltime by -\mins

\shorttitle{Ultrasmall Silicate Grains}

\begin{document}

\title{
        \vspace*{-2.0em}
        {\normalsize\rm submitted to {\it The Astrophysical Journal Letters}}\\
        \vspace*{1.0em}
On Ultrasmall Silicate Grains in the Diffuse Interstellar Medium\\
	}

\author{Aigen Li and B.T. Draine}
\affil{Princeton University Observatory, Peyton Hall,
        Princeton, NJ 08544, USA;\\
       {\sf agli@astro.princeton.edu, 
	draine@astro.princeton.edu}}

\begin{abstract}
The abundance of both amorphous and crystalline silicates 
in very small grains is limited by the fact 
that the 10$\mu$m silicate emission feature is not detected in 
the diffuse ISM. 
On the basis of the observed IR emission spectrum 
for the diffuse ISM, 
the observed ultraviolet extinction curve,
and the 10$\mu$m silicate absorption profile, we obtain
upper limits on the abundances of ultrasmall 
($a\ltsim15\Angstrom$) amorphous and crystalline 
silicate grains. 
Contrary to previous work, as much as $\sim$20\% of 
interstellar Si could be in $a\ltsim 15\Angstrom$ silicate grains
without violating observational constraints.
Not more than $\sim$5\% of the Si can be in crystalline silicates
(of any size).

\end{abstract}

\keywords{dust, extinction --- infrared: ISM: lines and bands 
--- ISM: abundances --- ultraviolet: ISM}

\section{Introduction}
The idea of interstellar silicate grains can be traced back to 
the 1960's when Kamijo (1963) first proposed that SiO$_2$, condensed 
in the atmospheres of cool stars and blown out into the interstellar
space, could provide condensation cores for the formation of ``dirty 
ices''. It was later shown by Gilman (1969) that grains around 
oxygen-rich cool giants are mainly silicates such as Al$_2$SiO$_3$ 
and Mg$_2$SiO$_4$. Silicates were first detected in emission in 
M stars (Woolf \& Ney 1969; Knacke et al.\ 1969), in the Trapezium 
region of the Orion Nebula (Stein \& Gillett 1969), and in comet 
Bennett 1969i (Maas, Ney, \& Woolf 1970); in absorption toward the 
Galactic Center (Hackwell, Gehrz, \& Woolf 1970), and toward the
Becklin-Neugebauer object and Kleinmann-Low Nebula (Gillett 
\& Forrest 1973). 
Silicates are now known to be ubiquitous, seen in 
interstellar clouds, circumstellar disks around young stellar objects
(YSOs), main-sequence stars and evolved stars, in HII regions, and in 
interplanetary and cometary dust.

The observed broad, featureless profile of the 10$\mu$m Si-O stretching mode
indicated that interstellar silicates are predominantly amorphous.
With the advent of {\it Infrared Space Observatory} (ISO) 
and the improved ground-based spectrometers, the presence of 
crystalline silicates has been revealed in six comets (see Hanner 1999
for a summary), in dust disks around main-sequence stars (see Artymowicz 
2000 for a summary), YSOs, and evolved stars (see Waelkens, Malfait, 
\& Waters 2000 for a summary), in interplanetary dust particles (IDPs) 
(Bradley et al.\ 1999), and probably also in the Orion Nebula (Cesarsky 
et al.\ 2000).

If silicate grains with radii $a\ltsim 15\Angstrom$ -- which we will
refer to as ``ultrasmall'' -- are present,
single-photon heating by starlight (Greenberg 1968) 
will cause these grains
to radiate in the 10$\micron$ feature
(see, e.g., Draine \& Anderson 1985).
The absence of a 10$\micron$ emission feature in IRAS spectra led
D\'esert et al.\ (1986) to conclude that not more than 1\% of Si could be
in $a < 15\Angstrom$ silicate grains, and on this basis recent grain models
(Duley, Jones, \& Williams 1989; D\'{e}sert, Boulanger, \& Puget 1990;
Siebenmorgen \& Kr\"{u}gel 1992; Mathis 1996; Li \& Greenberg 1997; 
Dwek et al.\ 1997; Weingartner \& Draine 2001; Li \& Draine 2001) 
have excluded ultrasmall silicate grains as a significant grain component.
Nondetection of 
the 10$\mu$m silicate emission feature 
in the diffuse ISM by ISO (Mattila et al.\ 1996)
and the {\it Infrared Telescope in Space} (IRTS) (Onaka et al.\ 1996) 
appeared to confirm this conclusion.

However, the presence of ultrasmall silicate grains can not be ruled out 
since the 10$\mu$m silicate emission feature may be hidden by 
the dominant PAH features. 
In this {\it Letter} we place quantitative upper limits on
the abundances of both amorphous and crystalline ultrasmall silicate 
grains by calculating spectra for such tiny grains heated
by starlight, and comparing to 
measurements of the infrared (IR) {\it emission}
of the diffuse ISM by 
IRTS (Onaka et al.\ 1996) and by
the {\it Diffuse Infrared Background 
Experiment} (DIRBE) instrument on the {\it Cosmic Background Explorer} 
(COBE) satellite.
The interstellar extinction curve and the 10$\mu$m silicate
{\it absorption} profile are also invoked to provide further 
constraints.

In \S\ref{sec:opct_si} we discuss the optical and thermal properties of 
crystalline silicate material. In \S\ref{sec:upplimt} we calculate the 
IR emission spectra of ultrasmall silicate grains and deduce upper limits 
on their abundances based on the constraints provided by the observed IR 
emission spectrum (\S\ref{sec:upplimt_emsn}), by the interstellar extinction 
curve (\S\ref{sec:upplimt_ext}), and by the 10$\mu$m silicate absorption 
profile (\S\ref{sec:upplimt_silabs}). In \S\ref{sec:discussion} we discuss 
the effects of dust size distributions and the sources of uncertainty. 
\S\ref{sec:summary} summarizes our main conclusions.

\section{Optical and Thermal Properties of Amorphous and 
Crystalline Silicates\label{sec:opct_si}}

We approximate the grains as spherical
and use Mie theory to calculate their absorption 
and scattering properties. 
The complex refractive index
${\rm m}(\lambda) = {\rm m}^\prime(\lambda) + 
i~ {\rm m}^{\prime\prime}(\lambda)$ of amorphous
silicates is taken from Draine \& Lee (1984).

For crystalline silicates, we proceed as follows:
for $\lambda \le 0.3\mu$m, we take ${\rm m}^{\prime\prime}$ from 
Huffman \& Stapp (1973) for crystalline olivine (Mg,Fe)$_2$SiO$_4$;
for $0.3< \lambda \le 6\micron$, we take ${\rm m}^{\prime\prime}$ from 
``astronomical silicate'' (Draine \& Lee 1984);
for $7 < \lambda \le 200\micron$, 
we take ${\rm m}^{\prime\prime}$ from Mukai \& Koike (1990) for crystalline 
olivine Mg$_{1.8}$Fe$_{0.2}$SiO$_{4}$; for $\lambda > 200\mu$m, 
we assume 
${\rm m}^{\prime\prime}(\lambda)={\rm m}^{\prime\prime}(200\mu {\rm m})
(200\micron/\lambda)^2$.
The real part ${\rm m}^\prime(\lambda)$ 
is obtained from ${\rm m}^{\prime\prime}(\lambda)$ 
using the Kramers-Kronig relation.

The specific heat of amorphous silicates is taken from 
Draine \& Li (2001). For crystalline silicates, following 
Draine \& Li (2001), we adopt a three-dimensional Debye model 
with $\Theta \approx 720$K which closely reproduces
the specific heat of crystalline Mg$_2$SiO$_4$ measured by 
Robie, Hemingway, \& Takei (1982).

\section{Upper Limits on Ultrasmall Silicate Grains\label{sec:upplimt}}

\subsection{IR Emission Spectra\label{sec:upplimt_emsn}}

We will now derive upper limits on the abundances of ultrasmall silicate 
grains in the diffuse ISM based on comparison of the {\it observed} 
IR emission spectrum with the {\it calculated} emission spectrum of 
ultrasmall silicate grains. For this purpose, we require a 
line-of-sight with:
1) a good quality mid-IR emission spectrum; and estimates of
2) the starlight intensity 
$\chi_{\rm MMP}$ (relative
to the Mathis, Mezger, \& Panagia [1983] radiation field), 
and 3) the hydrogen column density $N_{\rm H}$.

\begin{figure}[h]
\begin{center}
\epsfig{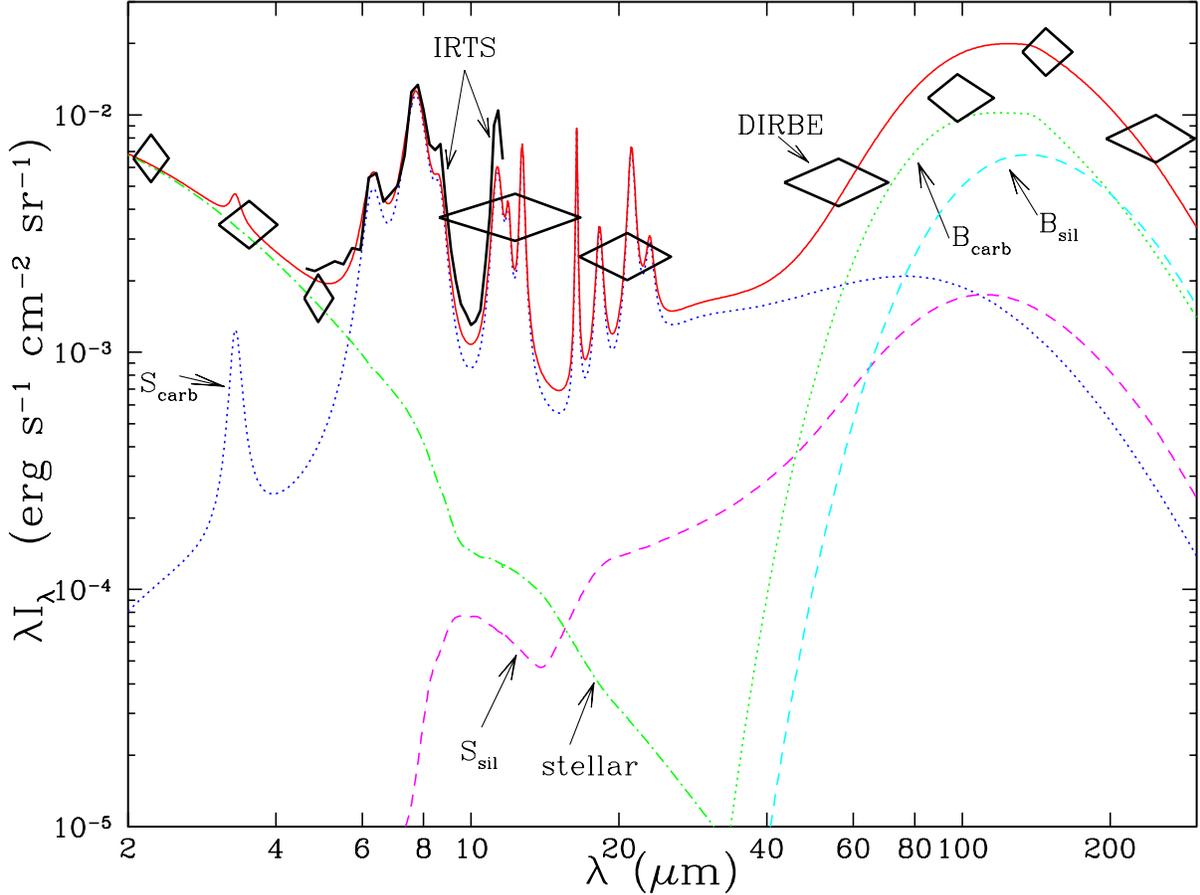}
\end{center}\vspace*{-1em}
\caption{
	\label{fig:MIRmodel}
	\footnotesize
	Spectrum toward $l=44^{\rm o}20^\prime$,
	$b=-0^{\rm o}20^\prime$ (see text). 
	Diamonds: DIRBE photometry 
	Heavy solid curve: MIRS/IRTS spectrum.
	Light solid curve: emission calculated for the
	carbonaceous/silicate grain model of 
	Li \& Draine (2001) heated by
	diffuse starlight with an intensity $\chi_{\rm MMP}=2$ relative 
	to the Mathis, Mezger, \& Panagia (1983) radiation field.
	Reddened starlight is labelled ``stellar''; curves labelled
	B$_{\rm carb}$,
	B$_{\rm sil}$ are contributions from ``big'' ($a > 250\Angstrom$)
	carbonaceous and silicate grains; S$_{\rm carb}$, S$_{\rm sil}$
	are contributions from ``small'' ($a < 250\Angstrom$) carbonaceous
	and silicate grains.
	}
\end{figure}

The diffuse ISM toward
$l = 44^{\rm o}20^\prime$, 
$b=-0^{\rm o}20^\prime$ has been observed
by 
DIRBE
(Hauser et al.\ 1998)
and the Mid-Infrared Spectrograph (MIRS) on 
IRTS 
(Onaka et al.\ 1996) has obtained the 4.7--11.7$\mu$m spectrum.
Li \& Draine (2001) have modelled the IR emission from the dust in this
direction. 
In Figure \ref{fig:MIRmodel} we show the observed photometry and mid-IR
spectrum 
together with the 
Li \& Draine (2001) model, with a total gas column 
$N_{\rm H}=4.3\times10^{22}\cm^{-2}$,
and a mixture of carbonaceous grains and silicate grains heated by
starlight with $\chi_{\rm MMP}\approx 2$.
This model is seen to be in good agreement with the
PAH emission features seen by IRTS, as well as the DIRBE photometry.
The total power radiated constrains the product 
$\chi_{\rm MMP}\times N_{\rm H}$,
while the $\sim$100$\micron$
wavelength of peak emission constrains $\chi_{\rm MMP}$.

Lacking {\it a priori} expectations concerning the form of the
size distribution for ultrasmall silicate grains,
we first consider grains of a single size $a$ 
(radius of an equal-volume sphere). 
Suppose a fraction $\fsi(a)$ of the total silicon abundance
(Si/H $\approx 36\times 10^{-6}$; Grevesse \& Sauval 1998)
is contained in amorphous silicate grains of radius $a$.
Applying the {\it thermal-discrete} method developed in
Draine \& Li (2001), we calculate the energy probability
distribution function and the resulting IR emission
spectrum.
The resulting emission spectra are shown in 
Figure \ref{fig:emsn_acsi}, where for each size $a$ the
value of $\fsi$ has been chosen such that this abundance of
ultrasmall amorphous silicate grains would not -- by itself --
exceed the observed emission at any wavelength.
Grains with $a \simlt10\Angstrom$  
are heated by energetic photons to ``temperatures'' high enough 
to emit in the 10$\micron$ silicate feature, and their abundances
are thus limited by the IRTS spectrum.
For grains with $a>10\Angstrom$ the emission peak shifts 
to longer wavelength, and the DIRBE 25$\micron$ data provides stronger
constraints.

From Figure \ref{fig:emsn_acsi}a
it can be seen that,
for $a < 10\Angstrom$, even if a fraction $\fsi=15\%$ 
were in the form of amorphous silicates, the resulting
10$\mu$m Si-O emission feature would not have been apparent in the IRTS
spectrum.
For $a=10\Angstrom$, $\fsi<25\%$ is permitted.
For $10< a < 20\Angstrom$, the DIRBE 25$\micron$ 
photometry allows up to $\fsi \simlt 30\%$,
and for $20<a< 25\Angstrom$, $\fsi\simlt 40\%$.

Similarly, we can estimate $\fsi(a)$ for ultrasmall crystalline
silicate grains. As shown in Figure \ref{fig:emsn_acsi}b, crystalline 
Mg$_{1.8}$Fe$_{0.2}$SiO$_{4}$ has pronounced features at 
10.0, 11.1, 16.3, 18.6, 23.0 
and 33.6$\mu$m. 
For $a < 10\Angstrom$, 
nondetection of the 10.0$\micron$ feature implies 
$\fsi \simlt 7\%$.
For $10 < a < 20\Angstrom$, an upper limit 
$\fsi \simlt 15\%$ 
is imposed by the DIRBE 25$\mu$m data,
and for $20<a<25\Angstrom$, $\fsi\ltsim20\%$.
Interstellar crystalline silicates could presumably differ
from the
crystalline olivine Mg$_{1.8}$Fe$_{0.2}$SiO$_{4}$
considered here,
resulting in changes in the strengths and positions of the 
infrared resonances (e.g., see J\"{a}ger et al.\ 1998 for 
different crystalline silicate samples).
Observations of comet Hale-Bopp (see Hanner 1999) showed an emission
feature at 11.2 rather than 11.1$\micron$, and an emission feature at
$10.0\micron$, evident in some spectra, was not always apparent.
Our limits on $a < 10\Angstrom$ crystalline silicates relied on the
strength and position of the 10.0$\micron$ feature; since this feature in
interstellar crystalline silicates might 
be either weaker or shifted in wavelength, it is appropriate to weaken
our upper limits for $a < 10\Angstrom$ to $\fsi \simlt 10\%$.
Below we will find that the 10$\micron$ absorption profile provides a
more restrictive limit on crystalline silicates.


\begin{figure}[h]
\begin{center}
\epsfig{
	file=f2.cps,
	width=\figwidth}
\end{center}\vspace*{-1em}
\caption{
	\footnotesize
        \label{fig:emsn_acsi}
        Contribution to the IR emission toward
        ($44^{\rm o}\le l \le 44^{\rm o}40^\prime$, 
        $-0^{\rm o}40^\prime\le b \le 0^{\rm o}$) by:
	(a) amorphous silicate grains with radii
	$a$=3.5\AA\ ($\fsi=15\%$),
	5\AA\ ($\fsi=15\%$),
	7.5\AA\ ($\fsi=15\%$),
	10\AA\ ($\fsi=25\%$),
	12.5\AA\ ($\fsi=30\%$),
	15\AA\ ($\fsi=30\%$),
	20\AA\ ($\fsi=30\%$),
	and 25\AA\ ($\fsi=40\%$);
        (b) crystalline silicate grains with radii
	$a$=3.5\AA\ ($\fsi=7\%$),
	5\AA\ ($\fsi=7\%$),
	7.5\AA\ ($\fsi=7\%$),
	10\AA\ ($\fsi=15\%$),
	12.5\AA\ ($\fsi=15\%$),
	15\AA\ ($\fsi=15\%$),
	20\AA\ ($\fsi=15\%$),
	25\AA\ ($\fsi=20\%$).
	Diamonds: DIRBE photometry.
	Heavy solid line:
        5--12$\micron$ spectrum observed by IRTS.
        }
\end{figure}

\subsection{Ultraviolet Extinction Curve\label{sec:upplimt_ext}}

\begin{figure}[h]
\begin{center}
\epsfig{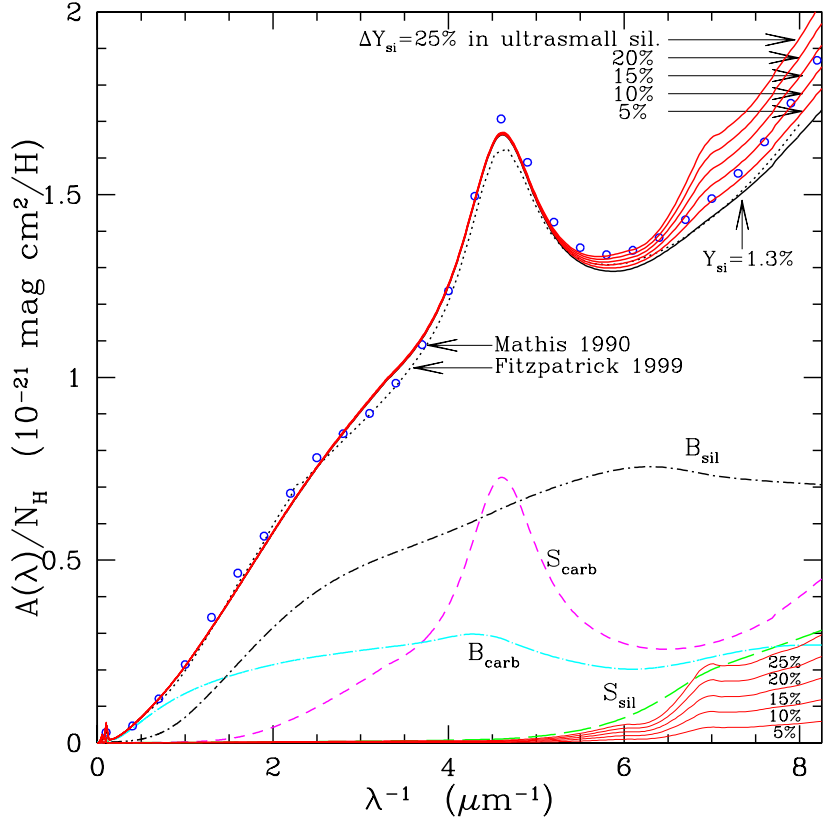}
\end{center}\vspace*{-1em}
\caption{
	\footnotesize
        \label{fig:ext_uv}
        Comparison of the average Galactic interstellar extinction 
        curve (open circles: Mathis 1990; dotted line: Fitzpatrick 
        1999) with theoretical extinction curves (solid lines)
        obtained by adding the contributions from ultrasmall silicate 
        grains with $\Delta\fsi$=5\%, 10\%, 15\%, 20\% and 25\% of the solar 
        silicon abundance (thin solid lines) to our best-fitting model 
        (with $\fsi=1.3\%$; 
	see Weingartner \& Draine 2001; Li \& Draine 2001). 
        Also plotted are the contributions (see Li \& Draine 2001) from 
        ``${\rm B_{sil}}$'' ($a \ge 250$\AA\ silicate);
        ``${\rm S_{sil}}$'' ($a < 250$\AA\ silicate); 
        ``${\rm B_{carb}}$'' ($a\ge 250$\AA\ carbonaceous); 
        ``${\rm S_{carb}}$'' ($a < 250$\AA\ carbonaceous, 
        including PAHs).          
        }
\end{figure}

Ultrasmall silicate grains 
make a steeply-rising contribution
to the ultraviolet (UV) extinction.
Weingartner \& Draine (2001) and Li \& Draine (2001) have obtained a
carbonaceous-silicate grain model which reproduces both
the observed extinction and the observed infrared emission.
This grain model has only a small abundance $\fsi\approx 1.3\%$ of 
ultrasmall ($a<15\Angstrom$) silicate grains.
In Figure \ref{fig:ext_uv}
we show the extinction obtained by {\it adding} ultrasmall 
silicate grains to this model.
Since the ultrasmall grains are
in the Rayleigh limit, the added extinction depends only on $\Delta\fsi$
(independent of grain size), and applies 
to both crystalline and amorphous silicates since their 
optical properties at $\lambda^{-1} \gtsim 5\micron^{-1}$ 
are taken to be similar (see \S\ref{sec:opct_si}).

As expected, the
contribution to the $\lambda^{-1}\simgt 7\mu {\rm m}^{-1}$ region
would be considerable for $\Delta\fsi\gtsim 20$\%
(although
the effects on the optical and near-UV extinction are 
negligible). 
In view of (1) the uncertainties in the observed interstellar 
extinction in this wavelength region (for example, the determinations 
of Mathis [1990] and Fitzpatrick [1999] differ from one other by 
$\sim 5\%$), 
(2) the fact that the size distributions 
of other components (carbonaceous grains and $a > 15\Angstrom$ silicates) 
could be adjusted
to reduce their contribution to the far-UV rise, 
and (3) the possibility that a mixture of ultrasmall silicates
(both neutral and charged) might have a more gradual onset of UV absorption
than our adopted silicate dielectric function
(which has a
relatively sharp increase in absorption near 6.5--7.0$\micron^{-1}$),
it seems clear that
the observed UV extinction does not rule out
$\Delta\fsi$ as large as 
$\sim 20\%$ on lines of sight with $R_V=3.1$.

\subsection{Silicate Absorption Profiles\label{sec:upplimt_silabs}}

\begin{figure}[h]
\begin{center}
\epsfig{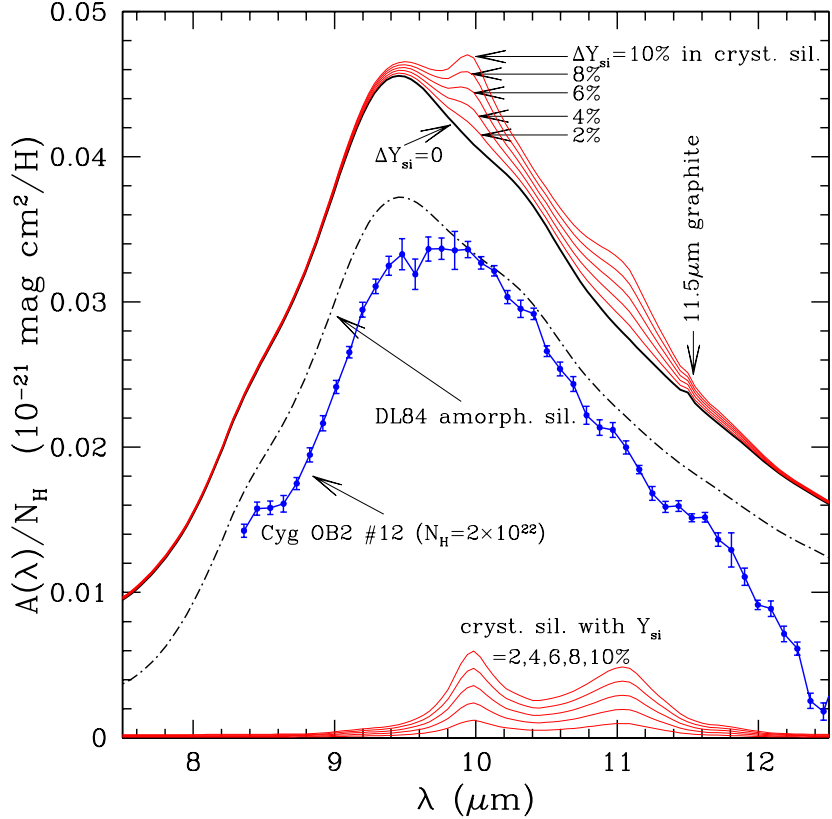}
\end{center}\vspace*{-1em}
\caption{
        \label{fig:ext_ir}
	\footnotesize
        The 10$\mu$m silicate absorption profile 
        (thick solid line) for the carbonaceous-amorphous silicate 
        model of Li \& Draine (2001)
        as well as profiles with an additional contribution from
        crystalline silicate grains with $\Delta\fsi$=2\%, 4\%, 6\%, 8\% 
        and 10\% of the solar silicon abundance (thin solid lines). 
        The narrow feature at 11.5$\mu$m is due to a lattice 
        mode of crystalline graphite (Draine 1984).
        Dot-dashed line is the absorption due to amorphous silicate
        (i.e., with the contributions of other dust components subtracted).
	Also shown is absorption which would be contributed by
	crystalline silicate grains for various $\fsi$.
	Also plotted is the observed extinction profile for dust toward
	Cyg OB2 \#12 (Bowey et al.\ 1998) divided by 
	$N_{\rm H}=2\times10^{22}\cm^{-2}$ (Whittet et al.\ 1997).
        }
\end{figure}

The 10$\mu$m silicate {\it absorption}
feature provides
another constraint on the abundance of
crystalline silicate grains of any size.
A recent high signal-to-noise spectroscopic study of
lines of sight towards Cyg OB2 \# 12 (diffuse ISM), Elias 16 
(quiescent molecular cloud) and Elias 18 (embedded YSO) in the
Taurus dark cloud appears to rule out any narrow absorption features
between 8.4 and 12.5$\mu$m (Bowey et al.\ 1998). 
Figure \ref{fig:ext_ir} 
plots the 10$\mu$m silicate absorption profiles resulting from 
addition of various amounts ($\Delta\fsi$) of ultrasmall crystalline 
silicates to our best-fitting model (see Weingartner \& Draine 2001; 
Li \& Draine 2001). 
The absence of narrow
features near 10.0 and 11.1$\micron$
in the observed 8.4--12.5$\micron$ extinction 
(Bowey et al.\ 1998)
suggests an upper limit $\Delta\fsi \simlt 3\%$ 
for crystalline silicates. 
Note that this result is insensitive to size
as the grains are in the Rayleigh limit.

As noted above, the precise wavelengths and strengths of IR resonance
features in crystalline silicates vary with composition, and there could be
more than one crystalline form present in
the interstellar medium.
Nevertheless, the lack of fine structure in the observed broad 10$\micron$
absorption feature makes it
unlikely that more than $\sim 5\%$ of the interstellar
Si could be in crystalline silicates.
This limit on crystalline silicates is more stringent than the
limit on ultrasmall crystalline silicates obtained in \S\ref{sec:upplimt_emsn}.

\section{Discussion\label{sec:discussion}}

The upper limits on the abundances of ultrasmall silicate
grains estimated from the IR emission spectra in \S\ref{sec:upplimt_emsn} 
are for grains of a single size. 
Adopting a log-normal size distribution 
$dn/d\ln a \propto \exp\{-[{\rm ln}(a/a_0)]^2/(2\sigma^2)\}$, 
we have calculated emission spectra for various $a_0$ and $\sigma$.
For this distribution, $a^3 dn/d\ln a$ peaks at $a_p = a_0e^{3\sigma^2}$.
For each distribution, we determined the maximum value of $\fsi$ such that
the infrared emission did not exceed either the IRTS 4.7--11.7$\micron$
spectrum or the DIRBE photometry.
Results are shown in 
Table \ref{tab:si_abun}. It can be seen that $\fsi$ increases with
$a_p$:
for $3\simlt a_0\simlt6\Angstrom$ and $0.3\simlt \sigma \simlt0.6$, 
$\fsi$ varies in the range of 20\%--55\%.
We take $\fsi\ltsim20\%$ to be the limit on $a<15\Angstrom$ silicate grains.
 
\begin{table}[h]
\caption[]{Upper limits on the abundance $\fsi$ of ultrasmall amorphous 
silicate grains placed by the absence of the 10$\mu$m silicate
emission feature (see \S\ref{sec:upplimt_emsn}, 
\S\ref{sec:upplimt_silabs})\label{tab:si_abun}.}
\begin{center}
\begin{tabular}{c c c c | 
		c c c c | 
		c c c c | 
		c c c c}
\hline \hline
$a_0$ & $\sigma$ & $a_p$ & $\fsi$ & 
	$a_0$ & $\sigma$ & $a_p$ & $\fsi$ &
	$a_0$ & $\sigma$ & $a_p$ & $\fsi$ &
	$a_0$ & $\sigma$ & $a_p$ & $\fsi$ \\
\hline
3\AA& 0.3& 3.9\AA& 20\%& 
	4\AA& 0.3& 5.2\AA& 20\%& 
	5\AA& 0.3& 6.5\AA& 20\%&
	6\AA& 0.3& 7.9\AA& 20\%  \nl 
3\AA& 0.4 & 4.8\AA& 20\%  & 
	4\AA & 0.4 & 6.5\AA& 20\% & 
	5\AA & 0.4 & 8.1\AA& 22\% &
	6\AA & 0.4 & 9.7\AA& 30\% \nl
3\AA    & 0.5 & 6.4\AA& 20\%  & 
	4\AA  & 0.5 & 8.5\AA& 25\% & 
	5\AA   & 0.5 & 10.6\AA& 35\% &
	6\AA & 0.5 & 12.7\AA& 45\% \nl
3\AA    & 0.6 & 8.8\AA& 30\%  & 
	4\AA  & 0.6& 11.8\AA& 40\% & 
	5\AA   & 0.6 & 14.7\AA& 50\% &
	6\AA & 0.6 & 17.7\AA& 55\%\nl
\hline
\end{tabular}
\end{center}
\end{table}

One uncertainty in our upper limit on ultrasmall amorphous
silicates arises from uncertainty in the absorption cross section for
$1 \ltsim \lambda^{-1} \ltsim 4\micron^{-1}$ where
the interstellar starlight spectrum peaks.
We have assumed the ultrasmall silicate grains to be characterized by
the refractive index estimated for amorphous ``astronomical silicates''
(Draine \& Lee 1984; Weingartner \& Draine 2001).
We note that the visual absorptivity of crystalline olivine material
is sensitive to its Fe content. 
If -- as suggested by the low Fe content observed in some circumstellar 
environments (see Waelkens et al.\ 2000) and in comet Hale-Bopp 
(Crovisier et al.\ 1997) -- {\it crystalline} astronomical
silicate material is less absorbing 
than ``astronomical silicate'' (Draine \& Lee 1984), then
photon absorptions will be less frequent and the IR emission
correspondingly reduced, in which case the infrared emission limit
on ultrasmall silicates would be relaxed.
However,
for crystalline silicates, the strongest constraint comes from
lack of sharp absorption features near 10 and 11$\micron$, so our upper
limit is not affected by uncertainty
in the 1--4$\micron^{-1}$ absorption.

The observed 3.3, 6.2, 7.7, 8.6, and 11.3$\mu$m emission 
bands (Mattila et al.\ 1996; Onaka et al.\ 1996) 
require a significant fraction of carbon in polycyclic aromatic
hydrocarbon molecules with $a\ltsim15\Angstrom$
(the model of Li \& Draine 2001 requires $\approx 40$ carbon 
atoms per $10^6$ H, or $\sim$10\% of the total C abundance).
In contrast, D\'esert et al.\ (1986) argued that not more than $\sim$1\% of
the Si could be in silicate particles with radii $a < 15\Angstrom$.
Such a paucity of ultrasmall silicate grains
would contrast sharply with the apparent large abundance of small
carbonaceous particles.
The present study, however, finds that as much as 20\% of the Si
could be locked up in ultrasmall silicate particles without violating
any observational constraints.

We suspect that the analysis by
D\'esert et al.\ (based on 2--13$\micron$ observations of M82 and NGC 2023) 
underestimated the fraction of the absorbed starlight energy which is 
reradiated in the 6--12$\micron$ PAH features.  
The present study is based on
a line of sight with photometry extending to the
far-infrared (see Figure \ref{fig:MIRmodel}). 

\section{Summary\label{sec:summary}}

Upper limits on the abundances of ultrasmall amorphous and 
crystalline silicate grains are estimated based on 
observed IR emission spectra (\S\ref{sec:upplimt_emsn}), 
the interstellar extinction curve (\S\ref{sec:upplimt_ext}), 
and the 10$\mu$m silicate absorption profile (\S\ref{sec:upplimt_silabs}). 
Our principal results are:

\begin{enumerate}
\item For amorphous silicate grains, 
non-detection of the 10$\mu$m silicate emission 
feature imposes an upper limit 
on the fraction $\fsi$ of interstellar Si in 
very small silicate grains:
$\fsi \simlt 15\%$ for radii $a<10\Angstrom$, and
$\fsi \simlt 20\%$ for $a = 10\Angstrom$, 
$\fsi \simlt 30\%$ for $10<a<20\Angstrom$,
and $\fsi \simlt 40\%$ for $20<a<25\Angstrom$.
For a distribution of sizes,
$\fsi \ltsim 20\%$ in grains with $a < 15\Angstrom$.
These limits are much higher than the earlier estimate by 
D\'{e}sert et al.\ (1986) of $\fsi\simlt 1\%$ for $a < 15\Angstrom$.

\item
The observed UV extinction curve in diffuse regions may allow as much as
$\fsi \simlt 20\%$ in ultrasmall silicate grains.

\item
For crystalline silicate grains, the absence of fine structure in the
observed 10$\mu$m 
silicate absorption profile provides the strongest constraint, limiting 
the abundances of $a\ltsim 1\micron$ crystalline silicate grains 
to $\fsi \simlt 5\%$.
\end{enumerate}

Toward high galactic latitude ``cirrus'' both the radiation field 
and the hydrogen column density are known, but mid-IR spectroscopy
is generally unavailable. 
In the near future, the {\it Space Infrared Telescope 
Facility} (SIRTF) may be able to obtain 
5--40$\micron$ spectra of cirrus regions to
either detect silicate emission or obtain 
more rigorous upper limits on 
the abundances of very small silicate grains.

\acknowledgments
We thank L.J. Allamandola, J.M. Greenberg, and J.C. Weingartner
for helpful discussions. 
We thank D.C.B. Whittet for providing us with the 10$\micron$ extinction
toward Cyg OB2 \#12.
We thank R.H. Lupton for the availability 
of the SM plotting package. This research was supported in part by 
NASA grant NAG5-7030 and NSF grants AST-9619429 and AST-9988126.

\end{document}